\documentstyle[twoside,fleqn,espcrc2,psfig]{article}

\newcommand{\AmS}{{\protect\the\textfont2
  A\kern-.1667em\lower.5ex\hbox{M}\kern-.125emS}}

\newcommand{\be}{\begin{equation}}
\newcommand{\ee}{\end{equation}}
\newcommand{\ben}{\begin{eqnarray}}
\newcommand{\een}{\end{eqnarray}}
\newcommand{\nn}{\nonumber}

\def\simgt{\rlap{\lower 3.5 pt\hbox{$\mathchar \sim$}}\raise 1pt \hbox {$>$}}
\def\simlt{\rlap{\lower 3.5 pt\hbox{$\mathchar \sim$}}\raise 1pt \hbox {$<$}}

\newcommand{\ce}{{\it c}_{\it E}}
\newcommand{\cb}{{\it c}_{\it B}}
\newcommand{\lqcd}{{\Lambda}_{\rm QCD}}

\title{Relativistic heavy quarks on the lattice}

\author{Sinya~Aoki\address{Institute of Physics,
             University of Tsukuba,
             Tsukuba, Ibaraki 305-8571, Japan},
        Yoshinobu~Kuramashi\address{
             High Energy Accelerator Research Organization(KEK),
             Tsukuba, Ibaraki 305-0801, Japan} and
        Shin-ichi~Tominaga\address{
             Center for Computational Physics,
             University of Tsukuba,
             Tsukuba, Ibaraki 305-8577, Japan}}
       
\begin{document}

\begin{abstract}
We investigate feasibility of relativistic approaches to the
heavy quark physics on both anisotropic and isotropic
lattices. Our peturbative calculation reveals that the
anisotropic lattice is not theoretically adovantageous over
the isotropic one to control $m_Q a$ errors.
We instead propose a new relativistic approach to handle
heavy quarks on the isotropic lattice.
\end{abstract}

\maketitle

\section{INTRODUCTION}
 
Lattice QCD is expected to provide the opportunity
of precise evaluation for weak matrix elements associated with $B$ mesons
from first principles. The main obstacle on the way is the
systematic errors originating from large $m_b a$ errors.
At present this difficulty is avoided in most cases by employing the
nonrelativistic approaches\cite{nrqcd,fermilab}. 
On the other hand, the use of
anisotropic lattice\cite{aniso} has a fascinating feature to
allow us to take the continuum limit.
However, there exists a theoretical concern
whether or not $m_Q a_s$
errors could revive perturbatively or nonperturbatively
even after they are removed classically.
Unfortunately, our one-loop calculation of the quark self
energy strongly suggests that radiative
corrections allow the revival of $m_Q a_s$ errors.

Having found that the anisotropic lattice is not
theoretically advantageous over the isotropic one, 
we propose a new
relativistic way to deal with the heavy quarks on the
isotropic lattice.
Cutoff effects in the heavy quark system are discussed by applying
the on-shell improvement program\cite{sym,onshell} 
to the finite $m_Q a$ case. 
We show that a proper adjustment of four parameters 
in the quark action reduces the remaining cutoff effects to be
$O((a\Lambda_{\rm QCD})^2)$. We also demonstrate
a determination of the four parameters at the tree-level from the
on-shell quark-quark scattering amplitude. 

In this report we present the salient points in our work\cite{akt}.

\section{ANISOTROPIC LATTICE}

In order to obtain a generic form of the quark action
on the anisotropic lattice, let us make the
operator analysis according to the on-shell
improvement program\cite{sym,onshell}. 
Under the requirement of various symmetries on the lattice,
we find the following set of operators are allowed up to
dimension five:
\ben
{\rm dim.3:}&& {\cal O}_3(x)={\bar q}(x)q(x), \\ 
{\rm dim.4:}&& {\cal O}_{4a}(x)={\bar q}(x)\gamma_0 D_0 q(x),\\
            && {\cal O}_{4b}(x)=
                 \mbox{$\sum_i$} {\bar q}(x)\gamma_i D_iq(x), \\ 
{\rm dim.5:}&& {\cal O}_{5a}(x)={\bar q}(x)D_0^2 q(x),\\
            && {\cal O}_{5b}(x)=\mbox{$\sum_i$} {\bar q}(x) D_i^2 q(x), \\
            && {\cal O}_{5c}(x)=
                 i\mbox{$\sum_i$} {\bar q}(x)\sigma_{0i}F_{0i} q(x),\\
            && {\cal O}_{5d}(x)=
                 i\mbox{$\sum_{i,j}$} {\bar q}(x)\sigma_{ij}F_{ij} q(x),\\
            && {\cal O}_{5e}(x)=
                 \mbox{$\sum_i$} {\bar q}(x) [\gamma_0 D_0,\gamma_i D_i] q(x), 
\een  
where 
the subscript $0$ denotes the time component, while $i,j=1,2,3$
space components.
Two degree of freedom in the eight coefficients are absorbed
in the renormalization of the quark mass $Z_m$ and the wave
function $Z_q$. We also find ${\cal O}_{5a}$ and 
${\cal O}_{5e}$ are related to
other operators by using the eq. of motion and hence they
are redundant. The remaining four parameters have to be
tuned to remove $O(a_{t,s})$ discretization errors.
Afterall we obtain the following expression for a general
form of the quark action on the anisotropic lattice:
\[
S_q= a_t a_s^3\mbox{ $\sum_x$} {\bar q}(x)\left[ \gamma_0 D_0
+\nu \mbox{$\sum_i$} \gamma_i D_i +m_0 \right. 
\]
\[
-a_t r/2 ( D_0^2+\eta \mbox{$\sum_i$} D_i^2)-a_t ig r/4  
\left\{ \cb \eta \mbox{$\sum_{ij}$} \sigma_{ij} F_{ij}
\right.
\]
\be
\left.\left.
+\ce(1+\eta)\mbox{$\sum_i$} \sigma_{0i} F_{0i}
\right\}\right]q(x),
\label{eq:qaction_d}
\ee
where the Wilson parameter $r$ is taken arbitrary.

We can determine the four parameters $\nu$, $\eta$, 
$c_E$ and $c_B$ at the tree-level from the on-shell
quark-quark scattering amplitude.
The improvement condition is that
the parameters should
be chosen to remove $O(a_t)$ cutoff effects from the
scattering amplitude at the on-shell point.
This is an extension of the previous work\cite{c_sw} done on the
isotropic lattice to the anisotropic one. 
Comparing ${\bar u}(p^\prime)\gamma_\mu u(p)$ $(p^\prime\ne
p$) on the lattice with the continuum expression, we
obtain $\nu=\eta=c_E=c_B=1$.
On the other hand, the $\nu$ and $\eta$ parameters can be
also determined
from the quark propagator by requiring that it should
reproduce the form,
\ben
S_q(p)&=&
\frac{1}{Z_q}
\frac{ -i \mbox{$\sum_\mu$}\gamma_\mu p_\mu + m_R}
{\mbox{$\sum_\mu$} p_\mu^2 +m_R^2}\nn\\
&&+{\rm (no\; pole\; terms)}+O(a_{t,s}^2)
\label{eq:qprop_r}
\een
with proper choice for $Z_q$, $Z_m$, $\nu$ and $\eta$.
We find $Z_q^{-1/2}=1-r m_0 a_t/2$,
$Z_m=1-r m_0 a_t/2$ and
$\nu=\eta=1$ at the tree-level. 
It should be noticed that the Dirac spinor takes the
correct relativistic form only when eq.~(\ref{eq:qprop_r})
is satisfied.


\begin{figure}[t]
\centering{
\hskip -0.0cm
\psfig{file=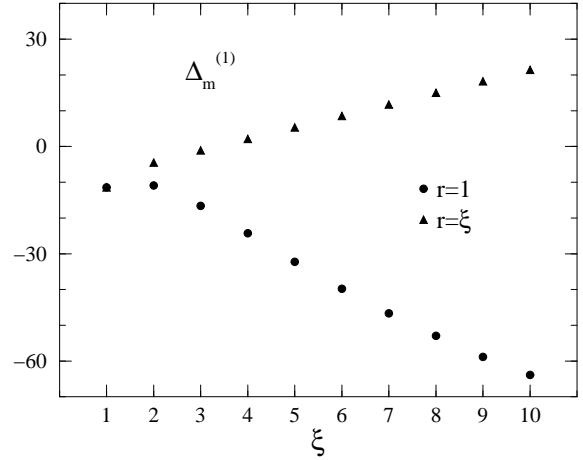,width=75mm,angle=-90}
\vskip -10mm  }
\caption{$\xi$ dependence of $\Delta_m^{(1)}$ in $Z_m$. Errors are
within symbols.} 
\label{fig:zm}
\vspace{-8mm}
\end{figure}

Expanding one-loop contributions to the quark self-energy
$\Sigma(p,m_0)$ in terms of $p$ and $m_0$,
we obtain the expressions of $Z_q^{-1/2}$, $Z_m$, $\nu$ and
$\eta$ up to $O(g^2 a_t)$.
As an example, the expression of $Z_m$ is given by
\[
Z_m=\left\{1+g^2 C_F/(16\pi^2) 
(-3{\rm log}(\lambda^2 a_s^2)+\Delta_m^{(0)})\right\}
\]
\be
\times\left\{1+a_t m\left(-r/2+g^2 C_F/(16\pi^2) \Delta_m^{(1)}
\right)\right\},
\ee
where $\lambda$ is fictitious gluon mass to regularize the
infrared divergence.
Figure~\ref{fig:zm} shows $\xi$ dependence of
$\Delta_m^{(1)}$. Here we consider $r=1$ and $r=\xi$ cases with
$\eta=1$, which satisfy the tree-level on-shell improvement condition.
We observe approximately linear dependences on $\xi$ for 
$\Delta_m^{(1)}$, which tells us that $O(g^2 a)$
contributions to $Z_m$ are effectively of order 
$g^2 m a_s=g^2m a_t\xi$. This leads us to conclude that
$m a_s$ errors are allowed to revive at the one-loop
level.
This is a reasonable conclusion in view of the on-shell
improvement. As far as we know there is no symmetry on the
anisotropic lattice to prohibit the higher dimensional operators
multiplied by $(m a_s)^n$.
Unless such symmetry is uncovered, the theoretical advantage of
the anisotropic lattice over the isotropic one would 
never be confirmed.

\section{ISOTROPIC LATTICE}

Let us explain our new relativistic approach to control 
$m_Q a$ errors for the heavy quarks on the isotropic lattice.
The basic idea is an application of the on-shell improvement
program to the finite $m_Q a$ case.
Here we assume that the leading cutoff effects are
$f_i(m_Q a,g^2,\log a)(a\lqcd)^i$ ($i\geq 0$), 
where $f_i(m_Q a,g^2,\log a)(a\lqcd)^i
>f_{i+1}(m_Q a,g^2,\log a)(a\lqcd)^{i+1}$ and 
$m_Q\gg \lqcd$.
$f_i(m_Q a,g^2,\log a)$  
are smooth and continuous 
all over the range of 
$m_Q a$ and have Taylor expansions at $m_Q a=0$ with
sufficiently large convergence radii beyond $m_Q a=1$.
We show that our formulation removes the cutoff effects
up to $f_1(m_Q a,g^2,\log a)a\lqcd$ with the use of the on-shell
improvement.
The remaining cutoff effects $f_2(m_Q a,g^2,\log a)(a\lqcd)^2$ could
be removed by continuum extrapolation or  may be small
enough to be neglected.

In Ref.\cite{akt},
we have derived the on-shell improved action from
the ``off-shell'' improved theory, which has the axis interchange 
symmetry.
Here we alternatively derive it without using
the axis interchange symmetry from the beginning.
In this case the discussion is exactly the same as for the
anisotropic case in the previous section.
The generic quark action at all order of $ma$ is given by
\[
S_q= a^4\mbox{ $\sum_x$} {\bar q}(x)\left[ \gamma_0 D_0
+\nu \mbox{$\sum_i$} \gamma_i D_i +m_0 \right. 
\]
\[
-a r_t/2 D_0^2-a r_s/2\mbox{$\sum_i$} D_i^2-a ig \cb/4  
\mbox{$\sum_{ij}$} \sigma_{ij} F_{ij}
\]
\be
\left.
-a ig \ce/2 \mbox{$\sum_i$} \sigma_{0i} F_{0i}
\right]q(x),
\label{eq:qaction_iso}
\ee
where we are allowed to choose $r_t=1$ and the four parameters
$\nu$, $r_s$, $c_E$ and $c_B$ are to be adjusted.
We remark that the parameter $r_s$ is not redundant, while
it is in Ref.\cite{fermilab}. 
As a consequence the ``$d_1$ field rotation'' required 
in Ref.\cite{fermilab} is not necessary in our formulation.
This difference originates from 
the ways to find the redundant operators:
we employ the classical field equation, while 
Ref.\cite{fermilab} uses the isospectral
transformation, which is originally proposed in
Ref.\cite{onshell}.
Although it is remarked in Ref.\cite{onshell} that this
transformation should be local and invariant under rotations
and reflections of the lattice,
Ref.\cite{fermilab} employs the
field transformation which breaks the rotational symmetry.
If the transformation is restricted to be rotationally
symmetric, which is actually proportional to the
classical field equation, the derived conclusion is exactly 
the same as ours.

The leading cutoff effects 
$f_0(m_Q a,g^2,\log a)$,
which reflect discretization errors of the 
operators  ${\bar q}(x)q(x)$,
${\bar q}(x)\gamma_0 D_0q(x)$ and ${\bar q}(x)D_0^2 q(x)$,
are completely absorbed in $Z_q$ and $Z_m$.
The next leading scaling violations stem from other terms 
in eq.(\ref{eq:qaction_iso}).
Their magnitude are expected to be
$f_1^{\rm W}(m_Q a,g^2,\log a)(a\lqcd)$
for the Wilson quark action and 
$f_1^{\rm SW}(m_Q a,g^2,\log a)(a\lqcd)$ with 
$f_1^{\rm SW}(m_Q a=0,g^2,\log a)=0$
for the (massless) $O(a)$ improved SW
quark action.
This implies that once we fix the pole mass from some spectral
quantity, the cutoff effects in other spectral quantities are
at most of order $a\Lambda_{\rm QCD}$, not $(m_Q a)^n$, 
for the Wilson and the
(massless) $O(a)$ improved SW quark actions.
If the four parameters in eq.(\ref{eq:qaction_iso}) are
properly adjusted, the remaining scaling violations are reduced to 
$f_2^{\rm ours}(m_Q a,g^2,\log a)(a\lqcd)^2$

The four parameters $\nu$, $r_s$, 
$c_E$ and $c_B$ at the tree-level are determined from
the on-shell quark-quark scattering amplitude in the same
way as for the anisotropic case.
The improvement condition that
the parameters should
be chosen to reproduce the continuum scattering
amplitude gives
$\nu={\rm sinh}(m_p)/m_p$, 
$r_s={\rm cosh}(m_p)/m_p+{\rm sinh}(m_p)(r_t-1/m_p)/m_p$, 
$c_E=r_t \nu$ and $c_B=r_s$.
Among these parameters, $\nu$ and $r_s$ 
can be determined by demanding that the tree-level
quark propagator should reproduce the continuum relativistic form. 
As expected, both methods give exactly the same values for
$\nu$ and $r_s$. 

\section{CONCLUSIONS}

The anisotropic lattice requires the four
parameters to be tuned for the $O(a)$ improvement 
even in the massless case, and allows the revival of 
$m_Q a_s$ errors through one-loop diagrams. 
As for the isotropic lattice our proposed action, which also
needs the adjustment of the four parameters, can reduce the
cutoff effects to be $O((a\lqcd)^2)$.

\vspace{3mm}

This work is supported in part by the Grants-in-Aid of the Ministry of 
Education (Nos. 13740169, 12014202, 12640253, 12304011,
13135204). 



\begin{thebibliography}{99}


\bibitem{nrqcd} 
G.~P.~Lepage {\it et al.},
Phys. Rev. D{\bf 46} (1992) 4052.

\bibitem{fermilab} A.~X.~El-Khadra, A.~S.~Kronfeld and P.~B.~Mackenzie,
Phys. Rev. D{\bf 55} (1997) 3933.

\bibitem{aniso} F.~Karsch,
Nucl. Phys. {\bf B205} (1982) 285;
G.~Burgers {\it et al.},
{\it ibid.} {\bf B304} (1988) 587.


\bibitem{sym} K.~Symanzik, 
Nucl. Phys. {\bf B226} (1983) 187, 205. 

\bibitem{onshell} M.~L\"{u}scher and P.~Weisz, 
Commun. Math. Phys. {\bf 97} (1985) 59. 

\bibitem{akt} S.~Aoki, Y.~Kuramashi and S.~Tominaga,
hep-lat/0107009.
   


\bibitem{c_sw} R.~Wohlert, 
DESY preprint 87-069 (1987), unpublished.

    









\end{thebibliography}
\end{document}